\documentclass[letterpaper, 10pt, conference]{ieeeconf}  

\usepackage{xcolor}

\definecolor{somegray}{rgb}{0.5, 0.5, 0.5} \newcommand{\darkgrayed}[1]{\textcolor{somegray}{#1}}
\makeatletter
\newcommand*\titleheader[1]{\gdef\@titleheader{#1}}
\AtBeginDocument{%
   \let\st@red@title\@title
   \def\@title{%
     \vskip-3em
     \bgroup\normalfont\large\centering\@titleheader\par\egroup
     \vskip1.5em\st@red@title}
}

\IEEEoverridecommandlockouts                              
\usepackage{multicol}
\usepackage[bookmarks=true]{hyperref}
\usepackage{subcaption}
\usepackage{cite}
\usepackage{acronym}
\usepackage{graphicx}
\usepackage{xcolor}
\usepackage{gensymb}
\usepackage{float}
\usepackage{amsmath,amssymb,amsfonts}
\usepackage{tabularx}
\usepackage{longtable}
\usepackage{booktabs}
\usepackage{multirow}
\usepackage{mathtools}
\usepackage[utf8]{inputenc}
\usepackage{color}
\title{\LARGE \bf
Event-driven Vision and Control for UAVs on a Neuromorphic Chip}

\titleheader{\darkgrayed{This paper has been accepted for publication at the\\ IEEE International Conference on Robotics and Automation (ICRA), Xi'an, 2021, \copyright IEEE}}

\author{Antonio Vitale$^{1}$, Alpha Renner$^{2}$, Celine Nauer$^{2}$, Davide Scaramuzza$^{3}$, and Yulia Sandamirskaya$^{1}$
\thanks{$^{1}$Antonio Vitale and Yulia Sandamirskaya are with Neuromorphic Computing Lab of  Intel, Intel Labs, Munich, Germany
        {\tt\small antonio.vitale@outlook.de, yulia.samdamirskaya@intel.com}}%
\thanks{$^{2}$Alpha Renner and Celine Nauer are with Institute of Neuroinformatics, University of Zurich and ETH Zurich, Switzerland
        {\tt\small alpren@ini.uzh.ch}}%
\thanks{$^{2}$Davide Scaramuzza is with the Robotic Perception Group, at both Institutes of Informatics, University of Zurich, and Neuroinformatics, University of Zurich and ETH Zurich, Switzerland
        {\tt\small sdavide@ifi.uzh.ch}}%
}

\begin{document}

\maketitle
\thispagestyle{empty}
\pagestyle{empty}

\begin{abstract}

Event-based vision sensors achieve up to three orders of magnitude better speed vs. power consumption trade off in high-speed control of UAVs compared to conventional image sensors. Event-based cameras produce a sparse stream of events that can be processed more efficiently and with a lower latency than images, enabling ultra-fast vision-driven control. Here, we explore how an event-based vision algorithm can be implemented as a spiking neuronal network on a neuromorphic chip and used in a drone controller. We show how seamless integration of event-based perception on chip leads to even faster control rates and lower latency. In addition, we demonstrate how online adaptation of the SNN controller can be realised using on-chip learning. Our spiking neuronal network on chip is the first example of a neuromorphic vision-based controller solving a high-speed UAV control task. The excellent scalability of processing in neuromorphic hardware opens the possibility to solve more challenging visual tasks in the future and integrate  visual perception in fast control loops.


\end{abstract}

\section{INTRODUCTION}

Autonomous unmanned aerial vehicles (UAVs) can potentially solve many tasks, e.g., autonomous infrastructure monitoring for predictive maintenance, supporting search and rescue operations, or delivering goods to remote areas. To be deployed in dynamic and unpredictable real-world environments, UAVs require fast and efficient perception and control~\cite{loianno2018special,delmerico2019current,falanga2019fast,falanga2020dynamic}. It has been shown that event-based cameras---visual sensors with asynchronous pixels reporting local luminance change, inspired by biological retinas~\cite{mahowald1994silicon,Lichtsteiner2006}---lead to up to three orders of magnitude faster visual processing than conventional image-based vision sensors, while also consuming a few mW of power~\cite{gallego2019event,falanga2020dynamic}. Recently, it has been demonstrated how an event-based camera can enable tracking of a simple visual pattern with a constrained UAV at an angular velocity of $>$1000 degrees per second--- unreachable performance with image-based sensors with on-board computing~\cite{dimitrova2020towards}.   

Event-based cameras, such as the Dynamic Vision Sensor (DVS)~\cite{davis,finateu20205}, generate a stream of events from their pixels instead of conventional image-matrices. Neither conventional computer vision algorithms nor convolutional neural networks (CNNs), typically used for advanced visual perception today, are directly applicable to this type of signals~\cite{gallego2019event}. Thus, special event-based vision algorithms have been developed to extract task-relevant information from the stream of events~\cite{DAngelo2020,Afshar2019,liu2019event,akolkar2020real}. For neural-network processing,  event-images are usually created by accumulating events in conventional matrix structures based on fixed time intervals or a fixed number of events~\cite{de2019cnn,gehrig2019end,Rebecq19pami,maqueda2018event}. Special purpose accelerators have been proposed to make the processing of such DNNs efficient for UAV applications~\cite{palossi201964}. 

Neuromorphic hardware originates from the same line of research as the DVS and realizes brain-inspired computing principles on-chip, such as: neuronal-network based computing, asynchronous event-based communication of neuronal activation with ``spikes'',  temporal dynamics of the state of neurons, and efficient on-chip learning in form of local synaptic plasticity. Neuromorphic computing devices---such as Intel's neuromorphic research chip Loihi~\cite{davies2018loihi} used in this work and other systems~\cite{Furber2012,Qiao2015}---support event-based computation directly and enable low-power spiking neural network architectures with on-chip learning. 

Spiking Neural Networks (SNNs) have been previously explored for vision tasks required for UAV control~\cite{gehrig2020event}, using end-to-end learning and CNN to SNN conversion. Here, we show how an SNN for solving a visual line-tracking task can be implemented directly on neuromorphic hardware. We compare the performance of the neuromorphic controller with the state of the art event-based vision-based controller, described in ~\cite{dimitrova2020towards}. 

In our previous work, we have introduced both SNN-based proportional (P) controllers as well as proportional, integral, derivative (PID) controllers, all implemented in neuromorphic hardware~\cite{GlatzEtAl2019,StagstedEtAl2020_IROS,StagstedEtAl2020_RSS}. This work inspired a number of recent neuromorphic motor control architectures~\cite{zhao2020closed} and continued the long-standing research line of neuronally inspired motor control methods~\cite{dewolf2020nengo,perez2013neuro}. 

In this paper, we bring this work two steps further towards low-power event-based vision-driven UAV control in real-world tasks: (1) using the same network architecture as in ~\cite{StagstedEtAl2020_IROS}, we demonstrate how adaptation of the controller can be realised in closed loop using on-chip learning rules; (2) we implement a SNN on Loihi which generates state estimates directly from the incoming DVS event stream. We demonstrate improved latency and processing rate compared to the CPU implementation of the same algorithm.

In contrast to previous work on ultra-fast event-driven motor control~\cite{conradt2009pencil,delbruck2013robotic, dimitrova2020towards}, here the DVS events are input  to the neuronal cores of a neuromorphic chip using an address event representation (AER) interface~\cite{boahen1999throughput}, and processed directly by the SNN. This enables future research into more complex SNNs for efficient on-chip perception to solve different tasks required for autonomous navigation and mission control: target tracking and obstacle avoidance, visual odometry, and map formation. 

We benchmark different parts of the SNN architecture and of the overall prototype system and conclude that the fully neuromorphic perception combined with conventional PD control can bring out another order of magnitude improvement in speed compared to conventional control driven by event-based vision, while relying on the low and scalable power-consumption  of neuromorphic hardware.  

\section{METHODS}
\subsection{Hardware}
\subsubsection{Neuromorphic hardware (Loihi)}
In this work, Intel’s neuromorphic research chip Loihi~\cite{davies2018loihi} in the form of a Kapoho Bay device was integrated on a rotating dual-copter. Kapoho Bay features two Loihi chips with a total of 256 neuro-cores in which a total of 262,144 neurons and up to 260 million synapses can be implemented. Three embedded x86 processors are used for monitoring and I/O spike management. 

\subsubsection{Constrained UAV}
\begin{figure}[H]
\begin{minipage}[t]{0.28\linewidth}
    \includegraphics[width=\linewidth]{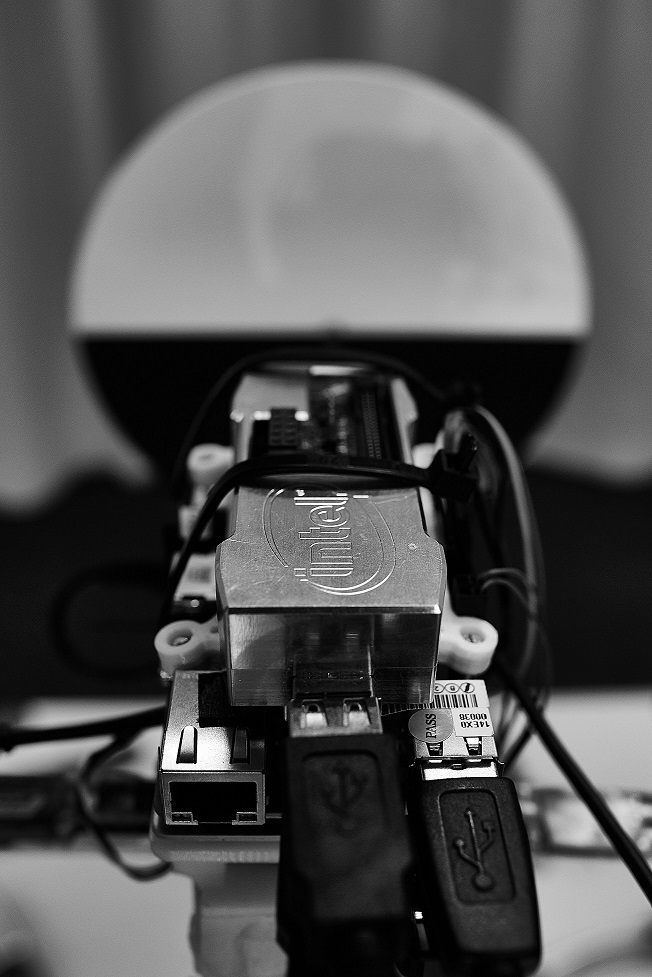}
\end{minipage}
\begin{minipage}[t]{0.63\linewidth}
    \includegraphics[width = \linewidth]{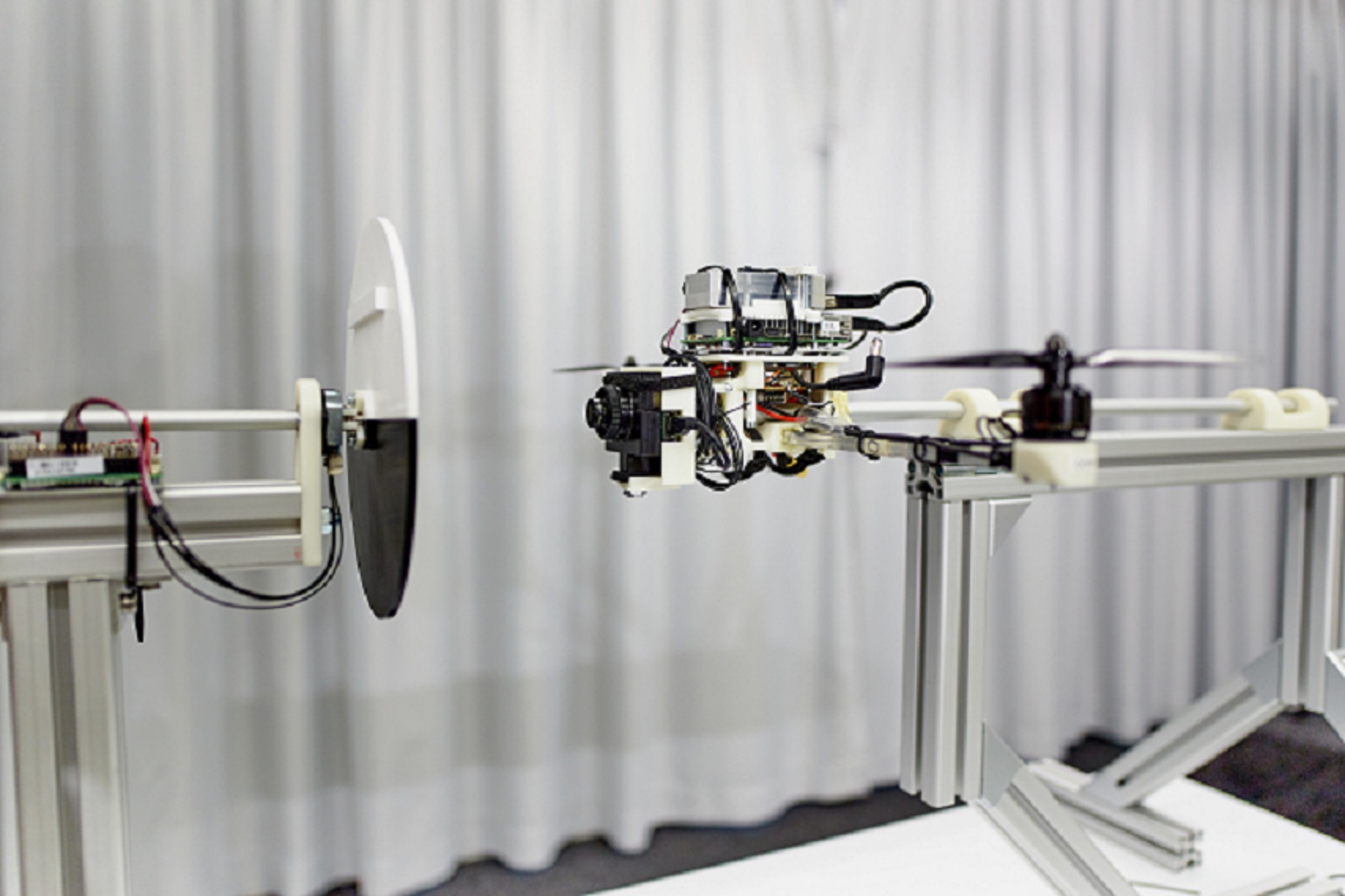}
\end{minipage}
\hfill
\caption{\textbf{Left}: Intel's Kapoho Bay mounted on top of the dual-copter. \textbf{Right}: Side view of the entire physical setup.}
\label{fig:test_bench}
\end{figure}

The same robotic platform was used as in \cite{dimitrova2020towards} - a constrained to 1 DoF dual-copter that can rotate 360 degrees in front of an actuated visual pattern.  A 3D-printed holder was designed to mount the Kapoho Bay on top of the rotating dual-copter. Fig.~\ref{fig:test_bench} shows both the mounted Kapoho Bay as well as a side view of the setup.
\subsubsection{Event Camera}
The dual-copter is equipped with a DAVIS 240C event camera with a spatial resolution of 240x180  pixels and a temporal resolution of the asynchronous event stream of 1$\mu$s~\cite{davis}. 

As in the previous work, the DAVIS sends the event stream to the UP board via USB 2.0. According to \cite{davis_time}, the duration of the vision processing, i.e., the elapsed time between an intensity change to the detection of the event by the on-board computer, is  $<$5ms.

In addition to this, we explore the possibilities of on-chip state estimation and control by connecting the DAVIS camera to the Kapoho Bay via a direct AER~\cite{boahen1999throughput} interface; the entire visual processing is done on the Loihi chip in this case.

\subsubsection{ESCs and Motors}

As in the original setup, the UP board sends motor commands to a Lumenier F4 AIO ﬂight controller using the universal asynchronous receiver/transmitter (UART) communication protocol with a maximum transmission rate of 4.1kHz.  The custom ﬁrmware for the ﬂight controller was used that enables communication with the motors’ electronic speed control (ESC) via the DShot150 protocol at a maximum rate of 9.4kHz. The ESC sends pulse-width modulation (PWM) signals to the motors, where the input voltage to the motors is given by the applied voltage (12.1V) times the duty cycle of the PWM signal.

The PD controller returns the required rotor thrusts to achieve the desired roll angle and these thrust values are converted to input voltages for the motors using a pre-calibrated thrust-to-duty-cycle mapping. The updated control commands are sent every $1ms$ to the flight controller.

\subsubsection{Rotary Encoders}

The dual-copter and disk are both equipped with CUI AMT22 Modular Absolute Encoders that measure the ground truth roll angle and velocity. The encoders have an accuracy of 0.2$\degree$ and a precision of 0.1$\degree$. The encoder transfers angle measurements to the UP board via Serial Peripheral Interface (SPI) with a maximum data transmission rate of almost 20kHz.

\subsubsection{Interfaces, system overview}
The neuromorphic device, Kapoho Bay, was embedded on an UP board featuring an Intel® ATOM™ x5-Z8350 Processor with 64 bits up to 1.92GHz and 4GB of RAM. The UP board was running Ubuntu 18.04, and the communication between the hardware was achieved with ROS Kinetic. Python 3.6.9 and version 0.9.9 of the Intel NxSDK were used to set up the SNN on the Loihi chip. All CPU results reported here were obtained on this UP board.\footnote{All performance measurements are based on testing as of October 2020 - March 2021 and may not reflect all publicly available security updates.}

The state variable used for the control loop is the difference in orientation between the drone and the line on the disk. We use two ways to obtain this measure: 1) As in the previous work, an event-based algorithm running on the UP board processes the event stream to obtain the state estimation; 2) In order to exploit the advantages brought by the integration of the neuromorphic sensor with the neuromorphic processor, we directly connect the DAVIS camera to the Kapoho Bay and run the visual processing on-chip.

\subsection{Spiking Neural Network (SNN) for Visual Processing}

The Hough transform is a geometry-based computer vision algorithm for line detection in images~\cite{Jiang_Bing_Huang_Knoll_2019,Duda_Hart_1972}. Points laying on the same line in the Cartesian space are mapped to intersecting sine curves in the Hough space (the $\theta$-$r$ space). Therefore, for many points on the same line, the $(\theta, r)$ pair with the highest intersection density represents the correct line in Cartesian space.

In~\cite{dimitrova2020towards}, an efficient version of the Hough transform for event-based input was used that updated an estimate of the Hough parameters of the line based on accumulated events every 1ms using a sliding 3ms window. The state estimation coming from the vision algorithm was updated at a rate of 1kHz to match the control command update rate, and a Kalman filter was used in order to smooth the state estimation.

Here, we propose a neuromorphic version of the Hough transform, similar to the previous computational work~\cite{seifozzakerini2018}. The incoming DVS events are integrated in a 2D array of neurons representing $(\theta, r)$. The incoming pixels at location $(x, y)$ are mapped to all  pairs of $(\theta, r)$ that parametrize a possible line intersecting the coordinate $(x, y)$. This effectively corresponds to doing the Hough transform by application of a single binary weight matrix to the incoming spikes. Due to the leaky integration of the neurons in the Hough layer, binning of spikes into event-frames is not necessary. 

DVS events are injected into neuronal cores in small batches at a fixed timestep duration. The state estimate is obtained at the same rate that correspond to the timestep duration. In our implementation, the computational timestep on the Loihi was $50\mu s$, thus allowing the state update rate of 20kHz, i.e. 20 times faster the required by the CPU controller. Instead of using a Kalman filter to smooth the estimate and to match the control command rate of 1kHz, a sliding average window with the size of 200 elements was applied to the raw signal at 20kHz and thus down-sampled to 1kHz. 

The binary connection matrix for the Hough transform is computed according equation $r = x\cdot\cos{\theta} + y\cdot\sin{\theta}$.
An excitatory synapse connects each pair of $(\theta, r)$ and $(x,y)$, for which the equation holds. The continuous values for $r$ are mapped on equally spaced bins of 10 pixels for $r \in (-200, 200)$, and the angles $\theta$ were mapped in the range $\theta \in (-90^{\circ}, 90^{\circ})$. Using a population of 90 neurons resulted in a resolution of $2^{\circ}$/neuron .

\begin{figure}[H]
\includegraphics[width= \linewidth,keepaspectratio]{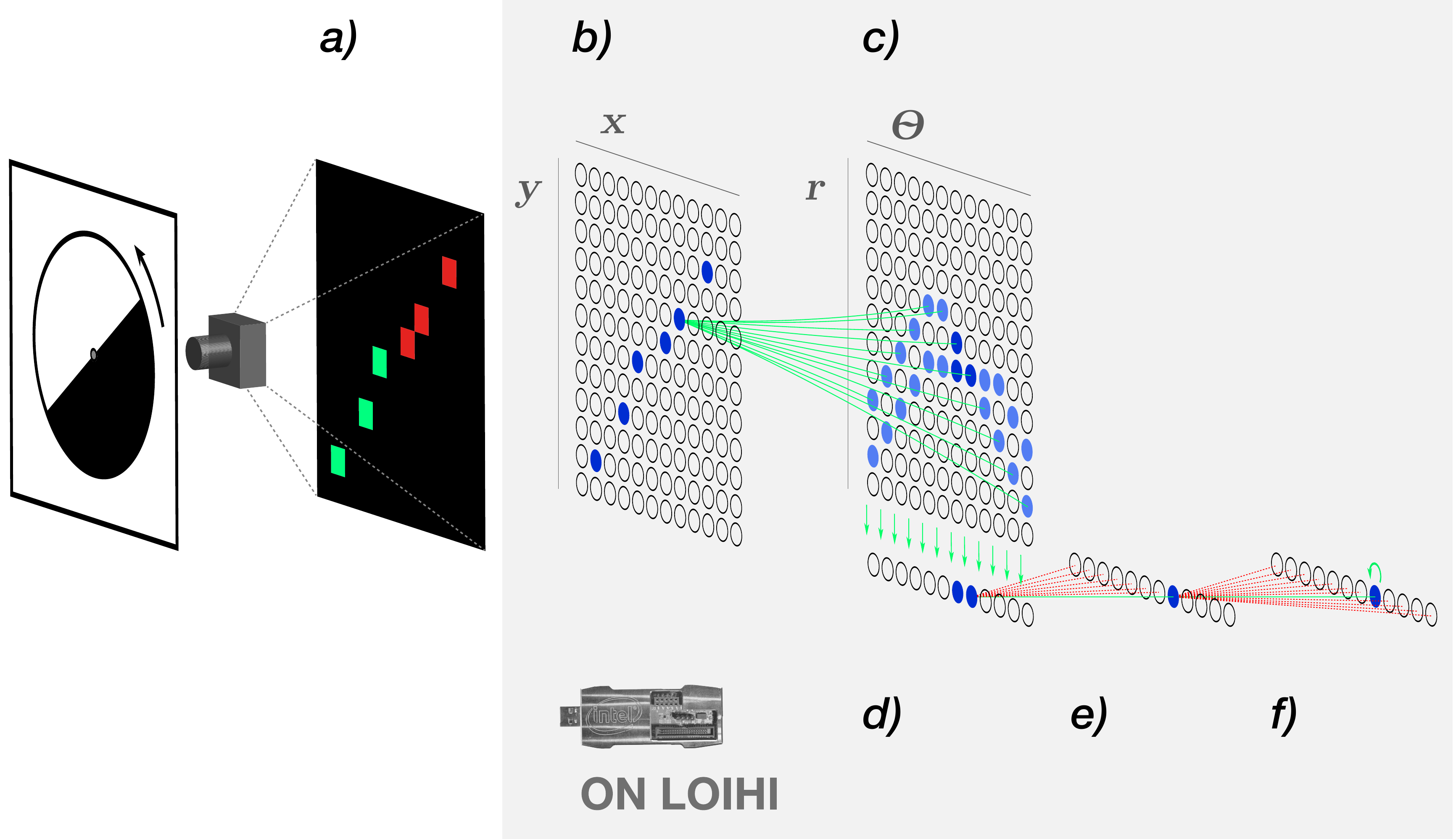}
\caption{On-Chip visual processing: a spiking neuronal network implementing the Hough transform on Loihi. The visual processing, from incoming events to estimated angle, on Loihi takes $250\mu s$. This corresponds to the time needed for events to be processed by the 5-layer network shown above with a computational timestep of $50\mu s$.}
\label{fig:visionsetup}
\end{figure}

The full SNN for the angular error estimation is illustrated in Fig.~\ref{fig:visionsetup}: the event-stream generated by the DVS camera (a) mounted on the drone pointing at a black-and-white pattern is down-sampled by a factor of 4 using an efficient 2-bit shift operation on the x86 co-processor on Loihi and then sent to the first layer of neurons (b) on the Loihi's neuro-cores. Neurons in layer (b) represent events in Cartesian Space of the camera frame; they are connected to neurons in layer (c) that represent events in the Hough Space -- according to the generated connectivity pattern. In order to accumulate events to generate an estimate for the current line orientation in the Hough space, the neurons in the 2D population (c) have a decay time of 3 timesteps and a threshold of $V_{threshold}=20 \cdot \mbox{Wgt}$, where $\mbox{Wgt}$ is the strength of the incoming weight, i.e. the update magnitude for each spike. This means that at least 20 events must lie on the same line within 3 timesteps, or $3 \cdot 50\mu s=150\mu s$, to generate an angle estimate.

In the layer (d) in Fig.~\ref{fig:visionsetup}, the 2D $(\theta, r)$ space is read out in $\theta$ direction by connecting all neurons that correspond to the same $\theta$ to a respective readout (``angle'') neuron, as seen in layer (d). These  neurons are then connected to a second population (e) of angle neurons that has the purpose of cleaning up activity in case several neurons are firing at the same time. 


The last of the three angle layers (f) acts as a memory due to self-excitatory connections storing the activity from the previous time step. This is needed in case the disk does not move and the event camera does not produce enough coherent spikes to bring the neurons in the Hough space above threshold. A new input, however, overwrites the stored angle by all-to-all inhibitory connectivity.

\subsection{The Adaptive SNN PD Controller} 
\subsubsection{Controller implementation}
The SNN used to control the drone's position is based on our previous work \cite{StagstedEtAl2020_IROS} and \cite{StagstedEtAl2020_RSS}, however, the on-chip SNN controller was simplified here: the two inputs to the controller are directly provided by the visual module (in CPU) as the angular error and its derivative. Also, in order to compare the CPU PD controller and the Loihi controller, only a PD controller was used, thus allowing to omit the whole integration pathway introduced in~\cite{StagstedEtAl2020_IROS}. Instead, we introduce an adaptive term in the controller that modifies the controller output in case the PD controller fails to remove the error (e.g., due to added load).

Since in our previous work we used a dual-copter moving at slower velocities and constrained to angles $\in (-25\degree, 25\degree)$, in addition to the faster computation time resulting from a better hardware integration, two major improvements were vital to obtain good performance on this faster platform:\\
1) More efficient allocation of neuron compartments on the neuro-cores allowed us to increase the population resolution and thus achieve more precise control and a larger range of commands. The neuron populations size was increased from $N=63$ to $N=361$ for the error and motor command populations. \\ 
2) The SNN output was obtained at a higher rate via a faster output communication channel between the neuromorphic processor and the host computer (UP board) using direct memory access instead of channel communication over the embedded CPU: the readout time decreased from $\approx 1.5ms$ to $\approx 0.05 ms$.  

The spikes in the output population of the controller $u$ encode the motor commands following the position encoding. Thus, the index of the firing neuron,  $idx \in[0,360]$, is decoded into the thrust difference according to Eq.~\ref{eq:output}. Here, $T_{min},T_{max} = \pm 1850$, and $N$ is the number of neurons in the output population:
\begin{equation}
\mathbf{u} =\frac{idx}{N} \cdot (T_{max} - T_{min}) + T_{min}.
\label{eq:output}
\end{equation}

The controller output is further converted in software to a thrust difference according to the equations:
\begin{equation}
\begin{split}
\mathbf{u} &= K_P \cdot \theta + K_D \cdot \dot\theta, \\
\mathbf{T} & = c_T \pm (\frac{u}{2} + b_T),
\label{eq:software}
\end{split}
\end{equation}
where $T$ is the thrust; the constant $c_T = 0.4 \cdot T_{baseline} = 2880$ ensures that the motors are always returning a baseline thrust. The thrust bias $b_T = 170$ is an empirically chosen value;  $u \in [-3700,3700]$ is the controller output. 



\subsubsection{Adaptation Pathway}
On-chip learning on Loihi enables adaption of the neuromorphic controller by using synaptic plasticity rules. Adaptive CPU controllers rely on monitoring the plant behavior and modifying the controller parameters accordingly.  On the Loihi chip, adaptation is realized by exploiting the plastic connections between compartments (neurons) directly on the processor, only minimally increasing the processing overhead. 


 To test this capability, we introduced a constant, static disturbance to the plant, i.e. added a  weight to one side of the drone.  To compensate for this a-priori unknown disturbance, we use an additive feed-forward layer that connects the input population $\theta$ to the output $u$. A similar approach was previously proposed for neuromorphic adaptive control of a robotic arm~\cite{dewolf2020nengo}.

The adaptive path should compensate for extended periods of large steady-state errors. Thus, it needs to monitor the error signal over time. If the accumulated error exceeds a threshold, then the feed-forward term should increase or decrease appropriately. The adaptive mechanism is regulated by the integral in Eq.~\eqref{equ:adap}. The monitoring period, $\Delta T$, and the cumulative error threshold, $\epsilon$, are parameters which are empirically selected:
\begin{equation}
\frac{1}{\Delta T} \int_{t}^{t+\Delta T} err^2(t) dt < \epsilon.
\label{equ:adap}
\end{equation}

\begin{figure}[H]
\includegraphics[width= \linewidth,keepaspectratio]{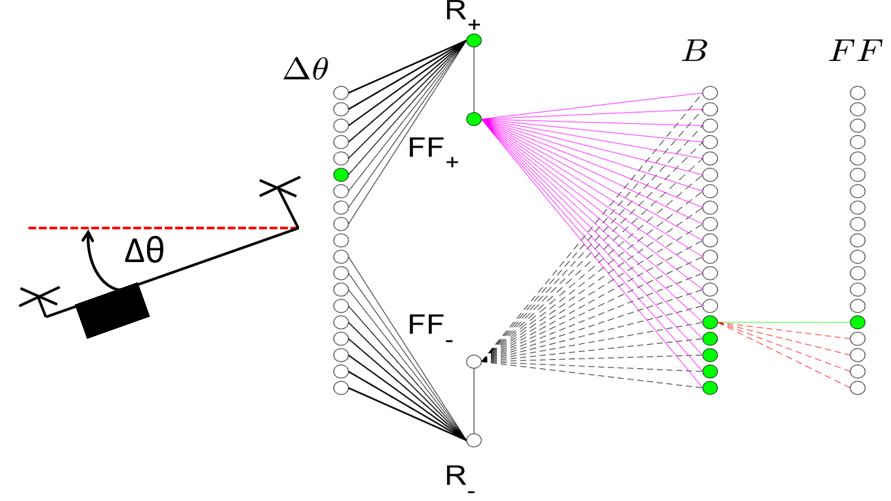}
\caption{Schematic of SNN Adaptation Pathway: Purple lines show currently active plastic synapses.}
\label{fig:snn_adap}
\end{figure}

In the network, we need to compute the integral in Eq.~\ref{equ:adap} using positional coding in the SNN. Fig.~\ref{fig:snn_adap} illustrates an equivalent SNN-pipeline: the neuronal population that represents the error signal, $err(t) = \Delta\theta$, is connected to two  neurons: one accumulating positive errors, $R_+$, and one -- negative errors, $R_-$. These neurons have the activation thresholds and decay constants such that they only fire if the accumulated error reaches a threshold value. Each neuron in the $\Delta \theta$ (error) population is connected to the $R_+$/$R_-$ neurons with a weight proportional to the error magnitude that the respective neuron represents: neurons closer to the center of the population, representing small errors, have a small weight and neurons on the periphery of the population (top and bottom in Fig.~\ref{fig:snn_adap}) are connected with larger weights. 

The R-neurons are connected to two intermediate neurons, $FF_+, FF_-$, which are respectively connected with positive and negative weights to the population $B$ and are thus responsible for increasing and decreasing the feed-forward term. The population $B$ uses a special form of position encoding: depending on the represented value (of the integrated error), more or less neurons are active in this population. To achieve this, neurons in $B$ have different activation thresholds (ordered from bottom to top in Fig.~\ref{fig:snn_adap}). 

Learning on Loihi is realised by local synaptic plasticity rules that can be programmed by the user. The learning rule can be composed by a combination of pre- and post-synaptic spikes, auxiliary variables representing recent activity, and a third factor, called the reinforcement channel. The R-neurons are connected to the reinforcement channel of the synapses between the $FF_+/FF_-$ neurons and the $B$ population in such a way that when the R neurons fire, the synaptic weight changes according to the learning rule $\Delta W = W \pm \delta_R(t)$ where $\delta_R(t)=1$ if the corresponding R neuron is firing, $\delta_R(t)=0$ otherwise.

Fig.~\ref{fig:snn_adap} shows the schematic of the adaptation pathway which runs in parallel to the full SNN PD controller. 

\subsection{SNN Input/Output Encoding/Decoding}
In order to send spikes to the Loihi chip, the host CPU (UP Board) communicates with one of the embedded processors of the Kapoho Bay using communication channels programmed with Intel's neuromorphic API NxSDK. To obtain the network output, the output spikes are sent to a FIFO file and read from the host computer.

When running the visual processing on Loihi, the output layer represents the angular offset between the drone and the disk. In order to keep the computation latency as small as possible, this angular offset was read out from the FIFO file on the host computer and then used as input to a PD controller running on the CPU. The current spiking index $\in (0,90)$ is linearly mapped to the angle range $\theta\in(-90\degree,+90\degree)$. For comparison: the control-loop running fully on CPU returns angle estimates $\theta\in(-180\degree,+180\degree)$ and uses $K_P=3000$ and $K_D=900$, while the control-loop running the visual processing on Loihi uses $K_P=2000$ and $K_D=600$ for angle estimates $\theta\in(-90\degree,+90\degree)$.

For the experiments done with the adaptive SNN Controller, where the vision-part of the controller was not the primary objective, the input to the SNN in Fig.~\ref{fig:snn_adap} is the  encoder on the drone's joint. In this case, the corresponding neurons receive an input spike from the on-chip spike generator according to the current roll angle of the drone.




\subsection{Experiments}
First, we demonstrate that the neuromorphic controller is able to perform with similar speed and accuracy as the state of the art event-driven CPU controller 
in an extreme control scenario, such as controlling a drone moving at high speeds. In order to maintain the disk at different constant speeds, a motorized wheel was attached to the axis of the disk and the rotational speed was kept constant for 10s. To ignore initial transient effects, the analysis of the data was done only for the second half (5s) of the experiment. To quantify the performance, the RMSE between the reading on the disk and drone encoders was calculated after correction for the delay, induced by the overall motor plant (on order of 100-150ms). 


To quantify the ability of the on-chip plasticity to adapt control parameters online, a reference trajectory was imposed on the drone to bring it to different target angles with an additional weight of 125g attached to one side of the drone's arm. The experiments were done once using the CPU PD controller and once using the adaptive SNN PD controller. In order to quantify the performance, we calculate the RMSE over a set of runs when using the adaptive SNN and the non-adaptive CPU controller.

\section{RESULTS}


\begin{figure}[h]
         \centering
         \includegraphics[width=\linewidth]{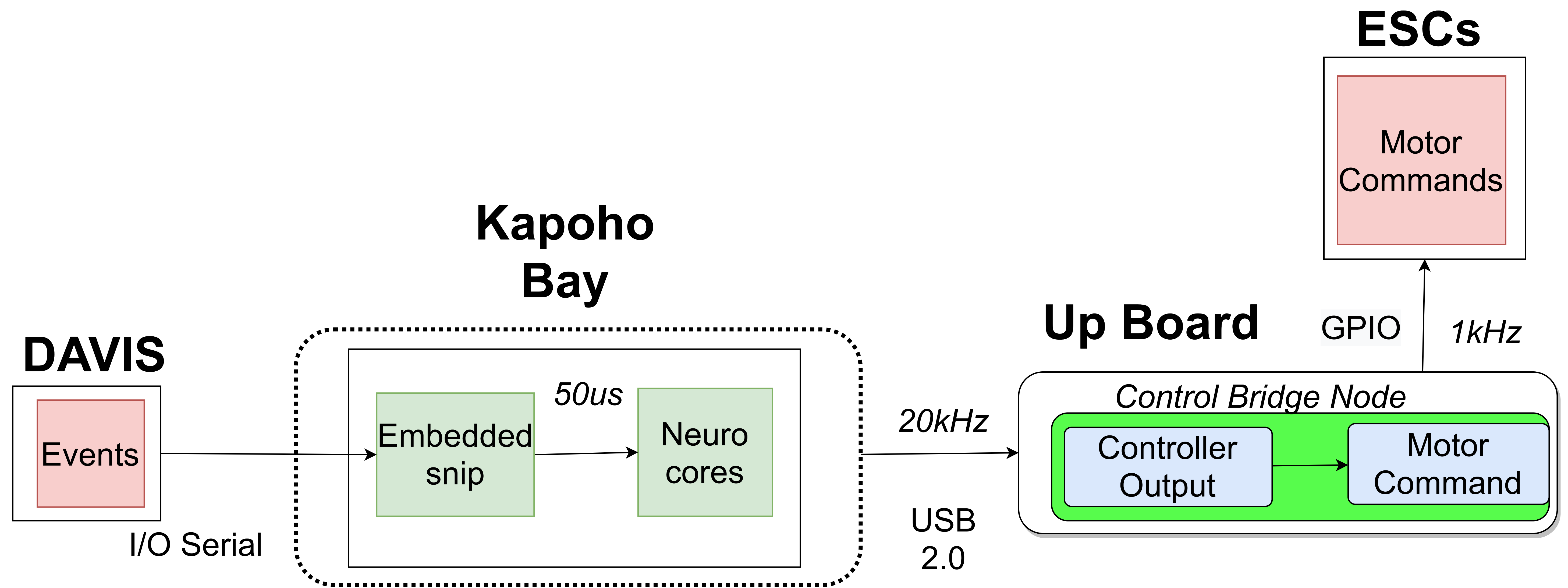}
        \label{fig:loi_lat}
\caption{The processing pipeline and vision-driven controller latency and update rate.}
\end{figure}

\subsection{Controller performance: High Speed Control}
\begin{figure}[H]
\includegraphics[width=\linewidth,keepaspectratio]{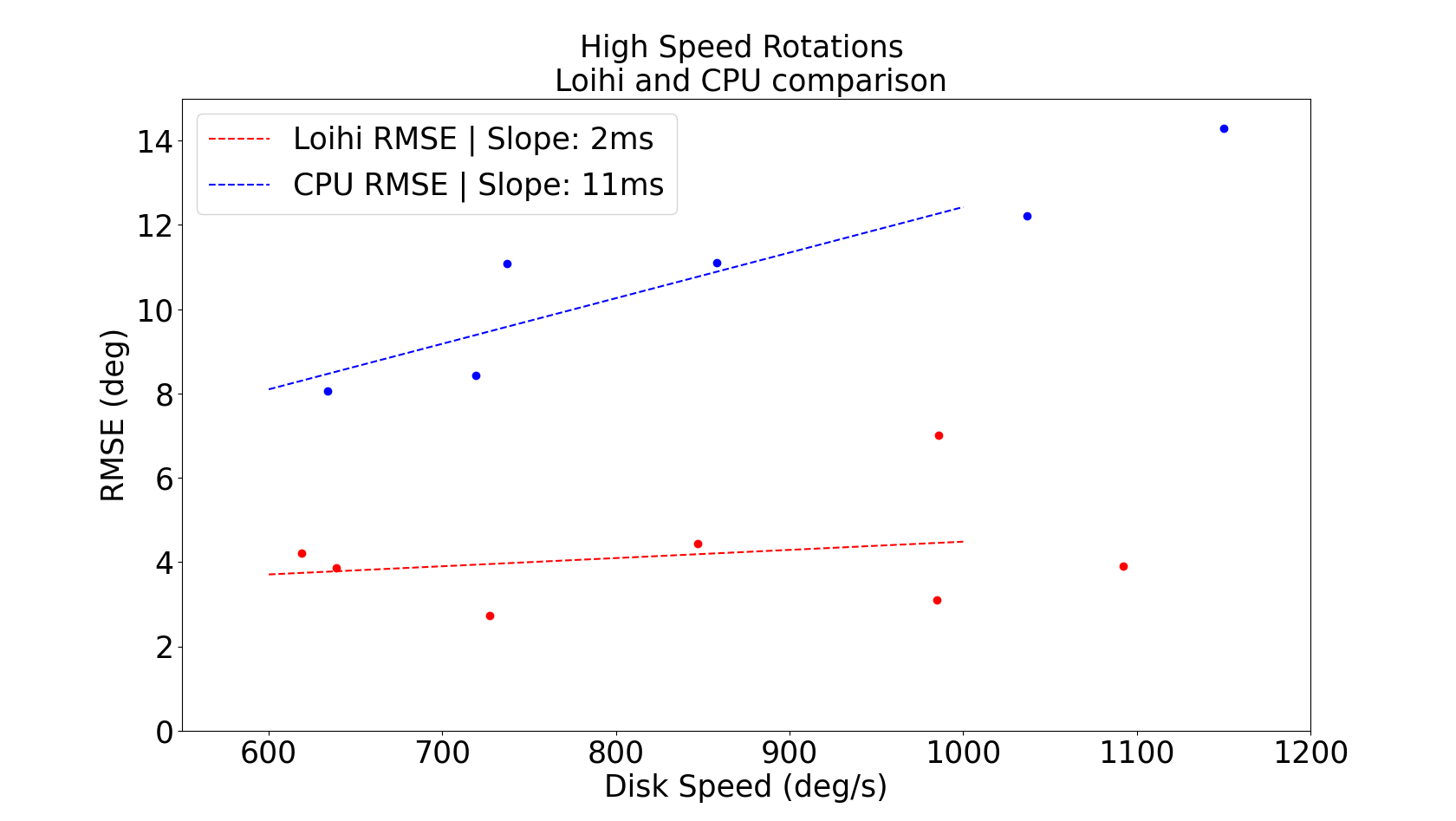}
\caption{RMSE when tracking the horizon using the CPU and Loihi visual processing at different angular velocities.}
\label{fig:rms}
\end{figure}

Here we test the ability of the drone to track the horizon rotating at different speeds. As expected, the controller which obtains the state estimation from Loihi can track higher speeds (of up to 1200 deg/sec) with a lower error than the controller with visual feedback computed in software running on the CPU. This increased performance relies on the visual processing running 20-times faster on Loihi than on the CPU (20kHz vs. 1kHz update rate). This allows for a more accurate estimate at high rotation speeds, ultimately leading to a slower error increase at raising speed, as can be seen from the plot in Fig.~\ref{fig:rms}. At lower speeds, we observed worse performance of the SNN-based angle estimation: the network's parameters were tuned for high speeds in our setup, hinting at the need for several controllers for different operation regimes.\footnote{The result at low rotation speed could not be obtained in the same motorized setup due to an  experimental limitation in disk actuation.}

\begin{figure}[h]
\includegraphics[width=\linewidth]{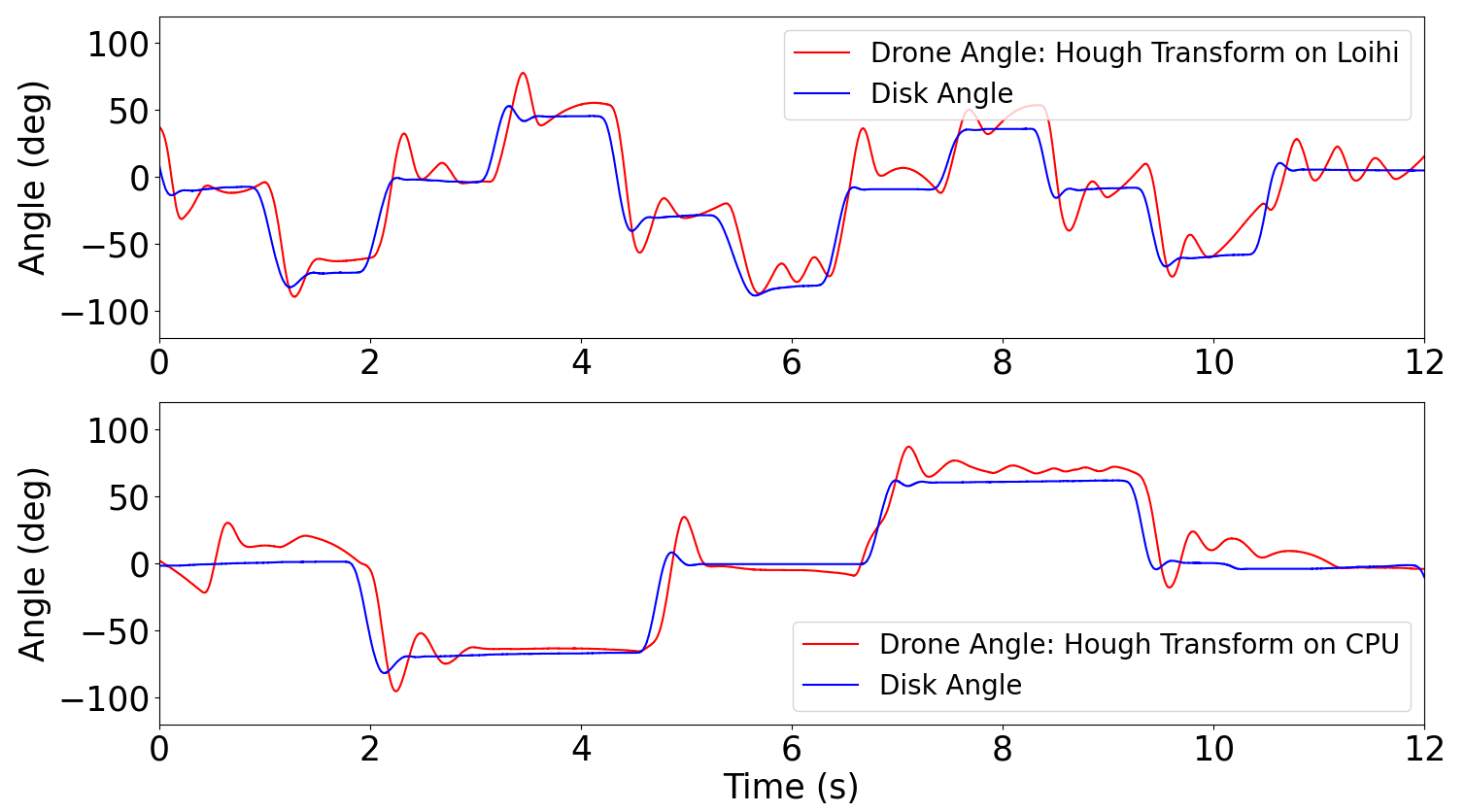}
\caption{\textbf{Top}: Line tracking using the Loihi visual processing as controller input. \textbf{Bottom}: Line tracking using the CPU visual processing as controller input.}
\label{fig:step}
\end{figure}

Fig.~\ref{fig:step} shows the closed-loop response when using the Loihi or the CPU to obtain an angle estimate from the visual feedback. In this experiment, the disk with the visual pattern was turned manually and the controller moved the drone to align it with the horizon line. One can observe that both controllers can solve the task.  The delay of the motor plant of the drone is on the order of 100-150ms, thus the speed of our SNN controller is under-exploited on this relatively slow system.

\subsection{Controller performance: Adaptive Control}
\begin{figure}[H]
\begin{minipage}[t]{\linewidth}
    \centering
    \includegraphics[width=0.8\linewidth]{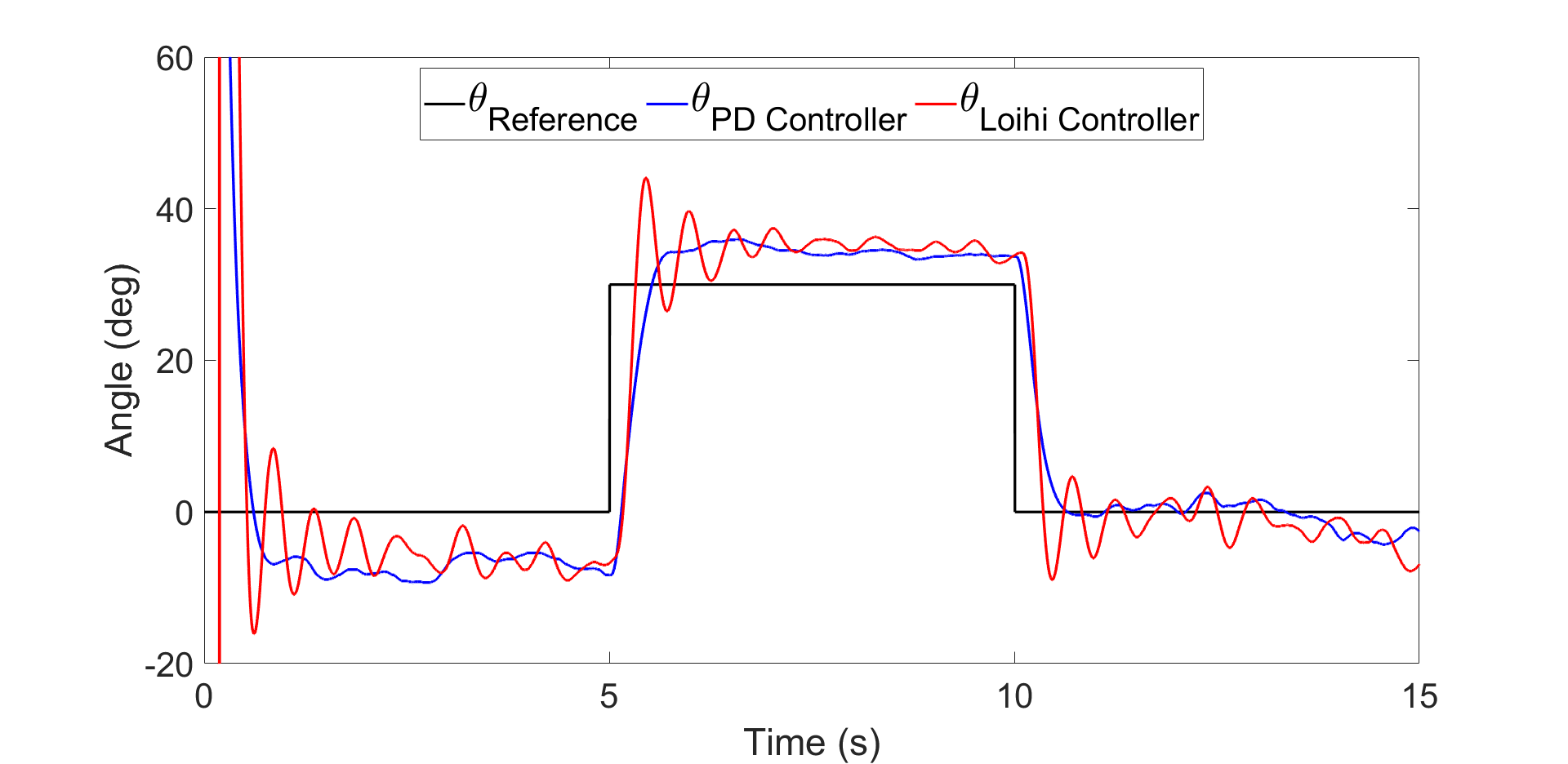}
\end{minipage}
\begin{minipage}[t]{\linewidth}
    \includegraphics[width = \linewidth]{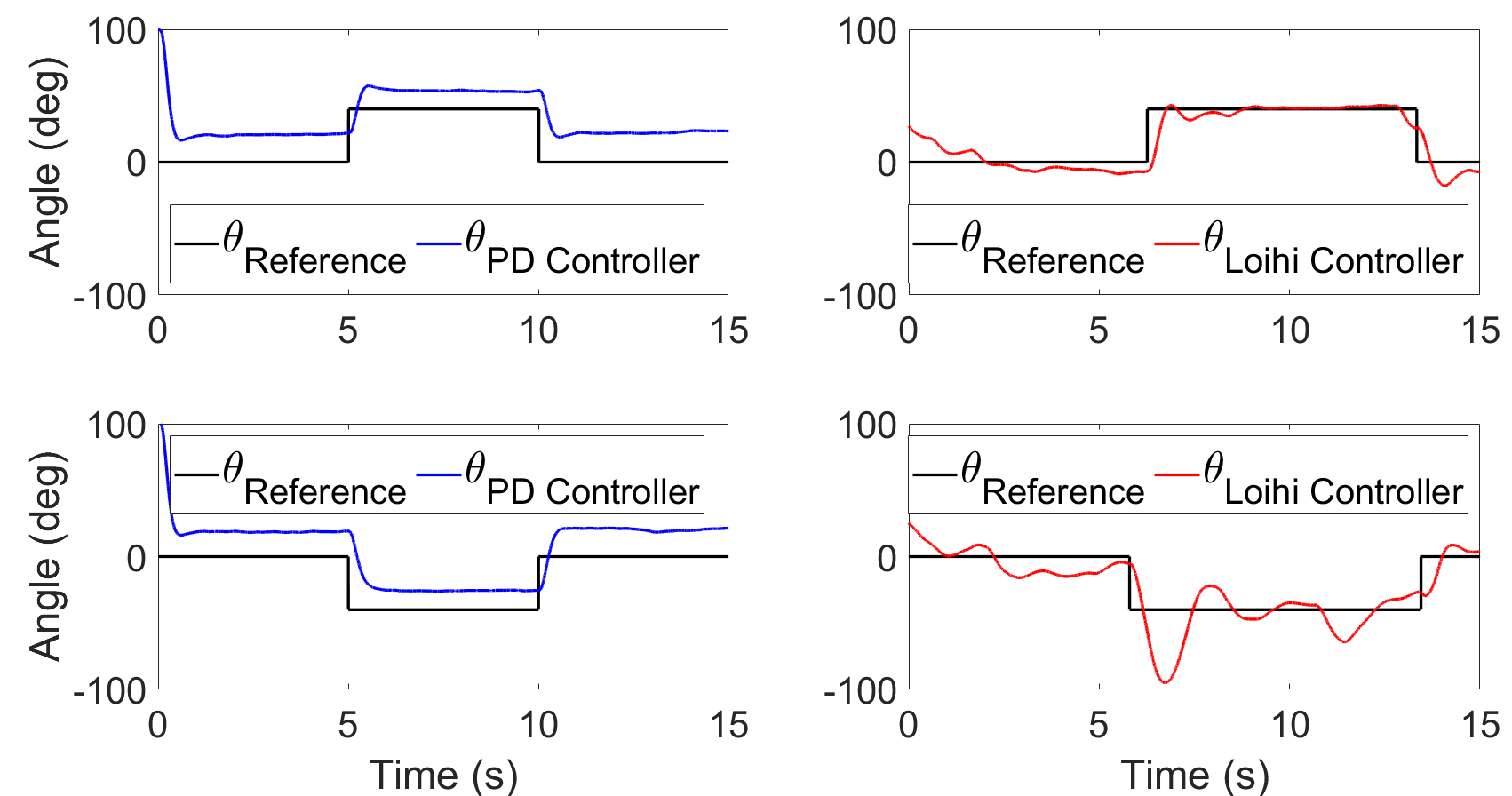}
\end{minipage}
\caption{Adaptive control experiments (without vision). \textbf{Top}: A trial run to illustrate the similar accuracy of both controllers. \textbf{Bottom}: \textbf{Left} Two trial runs using the software controller with the additional weight mounted to the drone, without adaptation. \textbf{Right} Same two trial runs using the adaptive SNN controller.}
\label{fig:enc_combine}
\end{figure}

In this series of experiments, we demonstrate the value of implementing the SNN controller in neuromorphic hardware, namely the possibility to explore on-chip learning to autonomously adjust the controller in teh closed loop. Fig.~\ref{fig:enc_combine} shows the performance of the CPU PD and SNN PD controller with adaptation when holding a set angular position using encoder feedback. Both controllers can solve this task: the obtained RMSE (Table~\ref{tab:encoder_track}) for the two controllers are similar, whereby a slightly increased error is noticeable for the experiments done with the SNN controller. This is due to the limited output resolution of the SNN controller (360 values) causing small oscillations around the set-points. The RMSEs were obtained by averaging over runs of 10 seconds each. 

The bottom plots in Fig.~\ref{fig:enc_combine} demonstrate how the adaptation,  implemented using synaptic plasticity on the neuromorphic chip, helps to remove the offset induced by an additional weight of 125g mounted on one of the UAV's arms. The plots show the roll of the drone at different target roll angles.  
 From the blue traces in Fig.~\ref{fig:enc_combine}, as well as from the obtained errors for the CPU controller reported in Table~\ref{tab:encoder_track}, it is clear that this additional weight of 125g represents a strong disturbance. Without adaptation, a persistent steady-state error of roughly $~20\degree$ is observed, while  the adaptation in the SNN controller removes this offset. With the added weight, the CPU controller's average error increases roughly by a factor of 2-3, while the adaptive controller manages to maintain the same error as without the disturbance. Note, the behavior of the adaptive controller here depends on the direction of the perturbation relative to the set point, showing the limitation of a simple single-value adaptive term. State-dependent adaptation can be explored in the SNN in future work. 

\begin{table}[h]
\caption{Accuracy of PD Controllers}
\scriptsize
\begin{tabular}[h]{*7c}
\toprule
Controller  & & & \textbf{RMSE (deg)}
\\
\midrule
\textit{{Step}} &\textit{+20$^{\circ}$}  & \textit{-20$^{\circ}$} & \textit{+30$^{\circ}$}  & \textit{-30$^{\circ}$}  & \textit{+40$^{\circ}$}  & \textit{-40$^{\circ}$} \\ 

\textbf{{No Added Weight}}
\\
Adaptive SNN PD&   6.9  & 12.3 & 8.2 & 13.2 & 11.4 & 13.9\\
Non-adaptive CPU PD&  6.9  & 7.7 & 8.5 & 9.2 & 9.4 & 11.5 \\
\hline

\textbf{{With Weight}}
\\
Adaptive SNN PD&   6.9  & 9.9 & 8.6 & 10.8 &8.5  & 16.0 \\
Non-adaptive CPU PD& 22.3  & 21.9 & 19.6 & 20.3 & 20.6 & 19.7 \\
\hline

\bottomrule
\end{tabular}
\label{tab:encoder_track}
\end{table}
\vspace{-5mm}










\section{CONCLUSIONS}
In this paper, we have shown  how neuromorphic vision-driven high-speed control can be implemented in practice. We presented the first neuromorphic controller with event-based visual feedback computed on a neuromorphic chip. The controller outperforms a state-of-the-art high-speed event-driven controller.

By integrating the event-based vision sensor more tightly with neuromorphic processors, we removed one of the bottlenecks in unleashing the full functionality of neuromorphic vision-driven control. The complexity of the visual task can be increased now by using large SNNs on chip, while maintaining lower latencies and lower power consumption, supported by good scaling properties of parallel and asynchronous neuromorphic hardware~\cite{davies2018loihi}. 

We also explored the possibilities of on-chip learning, enabled on Intel's Loihi chip,  when dealing with external disturbances. 

In the next iteration of our work, we aim to scale up the visual processing on the neuromorphic chip to solve more advanced vision tasks, such as place recognition or obstacle detection. We will further develop on-chip learning capability of the network to realize non-linear adaptive control.

\addtolength{\textheight}{0cm}   





\section*{ACKNOWLEDGMENT}
We thank Adrian Paeckel, Rebh\"usli Team for helping with the 3D printed parts; Thomas L\"angle and RPG team of UZH for support with the drone, and Intel's Neuromorphic Research Community for fruitful discussions and support. 

\bibliographystyle{IEEEtran}
\bibliography{literature}

\begin{thebibliography}{10}
\providecommand{\url}[1]{#1}
\csname url@samestyle\endcsname
\providecommand{\newblock}{\relax}
\providecommand{\bibinfo}[2]{#2}
\providecommand{\BIBentrySTDinterwordspacing}{\spaceskip=0pt\relax}
\providecommand{\BIBentryALTinterwordstretchfactor}{4}
\providecommand{\BIBentryALTinterwordspacing}{\spaceskip=\fontdimen2\font plus
\BIBentryALTinterwordstretchfactor\fontdimen3\font minus
  \fontdimen4\font\relax}
\providecommand{\BIBforeignlanguage}[2]{{%
\expandafter\ifx\csname l@#1\endcsname\relax
\typeout{** WARNING: IEEEtran.bst: No hyphenation pattern has been}%
\typeout{** loaded for the language `#1'. Using the pattern for}%
\typeout{** the default language instead.}%
\else
\language=\csname l@#1\endcsname
\fi
#2}}
\providecommand{\BIBdecl}{\relax}
\BIBdecl

\bibitem{loianno2018special}
G.~Loianno, D.~Scaramuzza, and V.~Kumar, ``Special issue on high-speed
  vision-based autonomous navigation of uavs,'' \emph{Journal of Field
  Robotics}, vol.~1, no.~1, pp. 1--3, 2018.

\bibitem{delmerico2019current}
J.~Delmerico, S.~Mintchev, A.~Giusti, B.~Gromov, K.~Melo, T.~Horvat, C.~Cadena,
  M.~Hutter, A.~Ijspeert, D.~Floreano \emph{et~al.}, ``The current state and
  future outlook of rescue robotics,'' \emph{Journal of Field Robotics},
  vol.~36, no.~7, pp. 1171--1191, 2019.

\bibitem{falanga2019fast}
D.~Falanga, S.~Kim, and D.~Scaramuzza, ``How fast is too fast? the role of
  perception latency in high-speed sense and avoid,'' \emph{IEEE Robotics and
  Automation Letters}, vol.~4, no.~2, pp. 1884--1891, 2019.

\bibitem{falanga2020dynamic}
D.~Falanga, K.~Kleber, and D.~Scaramuzza, ``Dynamic obstacle avoidance for
  quadrotors with event cameras,'' \emph{Science Robotics}, vol.~5, no.~40,
  2020.

\bibitem{mahowald1994silicon}
M.~Mahowald, ``The silicon retina,'' in \emph{An Analog VLSI System for
  Stereoscopic Vision}.\hskip 1em plus 0.5em minus 0.4em\relax Springer, 1994,
  pp. 4--65.

\bibitem{Lichtsteiner2006}
\BIBentryALTinterwordspacing
P.~Lichtsteiner, C.~Posch, and T.~Delbruck, ``{A 128 X 128 120db 30mw
  asynchronous vision sensor that responds to relative intensity change},''
  \emph{2006 IEEE International Solid State Circuits Conference - Digest of
  Technical Papers}, pp. 2004--2006, 2006. [Online]. Available:
  \url{http://siliconretina.ini.uzh.ch/wiki/lib/exe/fetch.php?media=lichtsteiner{\_}isscc2006{\_}d27{\_}09.pdf}
\BIBentrySTDinterwordspacing

\bibitem{gallego2019event}
G.~Gallego, T.~Delbruck, G.~Orchard, C.~Bartolozzi, B.~Taba, A.~Censi,
  S.~Leutenegger, A.~Davison, J.~Conradt, K.~Daniilidis \emph{et~al.},
  ``Event-based vision: A survey,'' \emph{IEEE Transactions on Pattern Analysis
  and Machine Intelligence (T-PAMI)}, 2020.

\bibitem{dimitrova2020towards}
R.~S. Dimitrova, M.~Gehrig, D.~Brescianini, and D.~Scaramuzza, ``Towards
  low-latency high-bandwidth control of quadrotors using event cameras,'' in
  \emph{2020 IEEE International Conference on Robotics and Automation
  (ICRA)}.\hskip 1em plus 0.5em minus 0.4em\relax IEEE, 2020, pp. 4294--4300.

\bibitem{davis}
C.~Brandli, R.~Berner, M.~Yang, S.-C. Liu, and T.~Delbruck, ``A 240x180 130db
  3$\mu$s latency global shutter spatiotemporal vision sensor,'' \emph{IEEE J.
  Solid-State Circuits}, vol.~49, no.~10, p. 23332341, 2014.

\bibitem{finateu20205}
T.~Finateu, A.~Niwa, D.~Matolin, K.~Tsuchimoto, A.~Mascheroni, E.~Reynaud,
  P.~Mostafalu, F.~Brady, L.~Chotard, F.~LeGoff \emph{et~al.}, ``5.10 a
  1280$\times$ 720 back-illuminated stacked temporal contrast event-based
  vision sensor with 4.86 $\mu$m pixels, 1.066 geps readout, programmable
  event-rate controller and compressive data-formatting pipeline,'' in
  \emph{2020 IEEE International Solid-State Circuits Conference-(ISSCC)}.\hskip
  1em plus 0.5em minus 0.4em\relax IEEE, 2020, pp. 112--114.

\bibitem{DAngelo2020}
G.~D'Angelo, E.~Janotte, T.~Schoepe, J.~O'Keeffe, M.~B. Milde, E.~Chicca, and
  C.~Bartolozzi, ``{Event-Based Eccentric Motion Detection Exploiting Time
  Difference Encoding},'' \emph{Frontiers in Neuroscience}, vol.~14, no. May,
  pp. 1--14, 2020.

\bibitem{Afshar2019}
S.~Afshar, T.~{Julia Hamilton}, J.~Tapson, A.~{Van Schaik}, and G.~Cohen,
  ``{Investigation of event-based surfaces for high-speed detection,
  unsupervised feature extraction, and object recognition},'' \emph{Frontiers
  in Neuroscience}, vol.~13, no. JAN, pp. 1--19, 2019.

\bibitem{liu2019event}
S.-C. Liu, B.~Rueckauer, E.~Ceolini, A.~Huber, and T.~Delbruck, ``Event-driven
  sensing for efficient perception: Vision and audition algorithms,''
  \emph{IEEE Signal Processing Magazine}, vol.~36, no.~6, pp. 29--37, 2019.

\bibitem{akolkar2020real}
H.~Akolkar, S.~H. Ieng, and R.~Benosman, ``Real-time high speed motion
  prediction using fast aperture-robust event-driven visual flow,'' \emph{IEEE
  Transactions on Pattern Analysis and Machine Intelligence}, 2020.

\bibitem{de2019cnn}
N.~De~Rita, A.~Aimar, and T.~Delbruck, ``Cnn-based object detection on low
  precision hardware: Racing car case study,'' in \emph{2019 IEEE Intelligent
  Vehicles Symposium (IV)}.\hskip 1em plus 0.5em minus 0.4em\relax IEEE, 2019,
  pp. 647--652.

\bibitem{gehrig2019end}
D.~Gehrig, A.~Loquercio, K.~G. Derpanis, and D.~Scaramuzza, ``End-to-end
  learning of representations for asynchronous event-based data,'' in
  \emph{Proceedings of the IEEE International Conference on Computer Vision},
  2019, pp. 5633--5643.

\bibitem{Rebecq19pami}
H.~Rebecq, R.~Ranftl, V.~Koltun, and D.~Scaramuzza, ``High speed and high
  dynamic range video with an event camera,'' \emph{IEEE Transactions on
  Pattern Analysis and Machine Intelligence}, 2019.

\bibitem{maqueda2018event}
A.~I. Maqueda, A.~Loquercio, G.~Gallego, N.~Garc{\'\i}a, and D.~Scaramuzza,
  ``Event-based vision meets deep learning on steering prediction for
  self-driving cars,'' in \emph{Proceedings of the IEEE Conference on Computer
  Vision and Pattern Recognition}, 2018, pp. 5419--5427.

\bibitem{palossi201964}
D.~Palossi, A.~Loquercio, F.~Conti, E.~Flamand, D.~Scaramuzza, and L.~Benini,
  ``A 64-mw dnn-based visual navigation engine for autonomous nano-drones,''
  \emph{IEEE Internet of Things Journal}, vol.~6, no.~5, pp. 8357--8371, 2019.

\bibitem{davies2018loihi}
M.~Davies, N.~Srinivasa, T.-H. Lin, G.~Chinya, Y.~Cao, S.~H. Choday, G.~Dimou,
  P.~Joshi, N.~Imam, S.~Jain \emph{et~al.}, ``Loihi: A neuromorphic manycore
  processor with on-chip learning,'' \emph{IEEE Micro}, vol.~38, no.~1, pp.
  82--99, 2018.

\bibitem{Furber2012}
S.~B. Furber, D.~R. Lester, L.~A. Plana, J.~D. Garside, E.~Painkras, S.~Temple,
  and A.~D. Brown, ``{Overview of the SpiNNaker System Architecture},''
  \emph{IEEE Transactions on Computers}, vol.~62, no.~12, pp. 2454--2467, 2012.

\bibitem{Qiao2015}
N.~Qiao, H.~Mostafa, F.~Corradi, M.~Osswald, F.~Stefanini, D.~Sumislawska, and
  G.~Indiveri, ``A reconfigurable on-line learning spiking neuromorphic
  processor comprising 256 neurons and 128k synapses,'' \emph{Frontiers in
  neuroscience}, vol.~9, p. 141, 2015.

\bibitem{gehrig2020event}
M.~Gehrig, S.~B. Shrestha, D.~Mouritzen, and D.~Scaramuzza, ``Event-based
  angular velocity regression with spiking networks,'' \emph{arXiv preprint
  arXiv:2003.02790}, 2020.

\bibitem{GlatzEtAl2019}
S.~Glatz, J.~Martel, R.~Kreiser, N.~Qiao, and Y.~Sandamirskaya, ``Adaptive
  motor control and learning in a spiking neural network realised on a
  mixed-signal neuromorphic processor,'' in \emph{2019 International Conference
  on Robotics and Automation (ICRA)}.\hskip 1em plus 0.5em minus 0.4em\relax
  IEEE, 2019, pp. 9631--9637.

\bibitem{StagstedEtAl2020_IROS}
R.~K. Stagsted, A.~Vitale, A.~Renner, L.~B. Larsen, A.~L. Christensen, and
  Y.~Sandamirskaya, ``Event-based pid controller fully realized in neuromorphic
  hardware: A one dof study.'' in \emph{IEEE/RSJ International Conference on
  Intelligent Robots and Systems (IROS)}, 2020.

\bibitem{StagstedEtAl2020_RSS}
R.~K. Stagsted, A.~Vitale, J.~Binz, A.~Renner, L.~B. Larsen, and
  Y.~Sandamirskaya, ``Towards neuromorphic control: A spiking neural network
  based pid controller for uav.'' in \emph{Robotics: Science and Systems},
  2020.

\bibitem{zhao2020closed}
J.~Zhao, N.~Risi, M.~Monforte, C.~Bartolozzi, G.~Indiveri, and E.~Donati,
  ``Closed-loop spiking control on a neuromorphic processor implemented on the
  icub,'' \emph{arXiv preprint arXiv:2009.09081}, 2020.

\bibitem{dewolf2020nengo}
T.~DeWolf, P.~Jaworski, and C.~Eliasmith, ``Nengo and low-power ai hardware for
  robust, embedded neurorobotics,'' \emph{arXiv preprint arXiv:2007.10227},
  2020.

\bibitem{perez2013neuro}
F.~Perez-Pe{\~n}a, A.~Morgado-Estevez, A.~Linares-Barranco,
  A.~Jimenez-Fernandez, F.~Gomez-Rodriguez, G.~Jimenez-Moreno, and
  J.~Lopez-Coronado, ``Neuro-inspired spike-based motion: from dynamic vision
  sensor to robot motor open-loop control through spike-vite,'' \emph{Sensors},
  vol.~13, no.~11, pp. 15\,805--15\,832, 2013.

\bibitem{conradt2009pencil}
J.~Conradt, M.~Cook, R.~Berner, P.~Lichtsteiner, R.~J. Douglas, and
  T.~Delbruck, ``A pencil balancing robot using a pair of aer dynamic vision
  sensors,'' in \emph{2009 IEEE International Symposium on Circuits and
  Systems}.\hskip 1em plus 0.5em minus 0.4em\relax IEEE, 2009, pp. 781--784.

\bibitem{delbruck2013robotic}
T.~Delbruck and M.~Lang, ``Robotic goalie with 3 ms reaction time at 4\% cpu
  load using event-based dynamic vision sensor,'' \emph{Frontiers in
  neuroscience}, vol.~7, p. 223, 2013.

\bibitem{boahen1999throughput}
K.~Boahen, ``A throughput-on-demand address-event transmitter for neuromorphic
  chips,'' in \emph{Proceedings 20th Anniversary Conference on Advanced
  Research in VLSI}.\hskip 1em plus 0.5em minus 0.4em\relax IEEE, 1999, pp.
  72--86.

\bibitem{davis_time}
E.~Mueggler, N.~Baumli, F.~Fontana, and D.~Scaramuzza, ``Towards evasive
  maneuvers with quadrotors using dynamic vision sensors,'' \emph{Eur. Conf.
  Mobile Robots (ECMR)}, p.~18, 2015.

\bibitem{Jiang_Bing_Huang_Knoll_2019}
Z.~Jiang, Z.~Bing, K.~Huang, and A.~Knoll, ``Retina-based pipe-like object
  tracking implemented through spiking neural network on a snake robot.''
  vol.~13, p.~29, May 2019.

\bibitem{Duda_Hart_1972}
R.~O. Duda and P.~E. Hart, ``Use of the hough transformation to detect lines
  and curves in pictures,'' \emph{Communications of the ACM}, vol.~15, no.~1,
  p. 11–15, Jan 1972.

\bibitem{seifozzakerini2018}
S.~Seifozzakerini, W.-Y. Yau, K.~Mao, and H.~Nejati, ``Hough transform
  implementation for event-based systems: Concepts and challenges.''
  \emph{Frontiers in Computational Neuroscience}, vol.~12, p. 103, Dec 2018.

\end{thebibliography}


\end{document}